\documentclass[
    reprint,
    amsmath, amssymb,
    aps,
    pra
    ]{revtex4-2}
\usepackage{graphicx}
\usepackage{dcolumn}
\usepackage[english]{babel}
\usepackage{bm}
\usepackage{xcolor}
\usepackage[utf8]{inputenc}
\usepackage{amsthm}
\usepackage{amsmath}
\usepackage{mathrsfs}
\usepackage{bm}
\usepackage{tikz-cd}
\usepackage{tikz}
\usepackage[toc,page]{appendix}
\usepackage{enumitem}
\usepackage{hyperref}

\usetikzlibrary{positioning}

\newcommand{\R}{\mathbb{R}}
\newcommand{\Z}{\mathbb{Z}}
\newcommand{\N}{\mathbb{N}}
\newcommand{\I}{\mathbb{I}}
\newcommand{\C}{\mathbb{C}}

\begin{document}
\preprint{APS/123-QED}
\title{Rapid-cycle Thouless pumping in a one-dimensional optical lattice}
\author{K.J.M. Schouten}\email{koen.schouten@student.uva.nl}
\author{V. Cheianov}

\affiliation{Insituut-Lorentz, Universiteit Leiden, Leiden, The Netherlands}

\date{\today}

\begin{abstract}
    An adiabatic cycle around a degeneracy point in the parameter space of a one-dimensional band insulator is known to result in an integer valued noiseless particle transport in the thermodynamic limit. 
    Recently, it was shown that in the case of 
    an infinite bipartite lattice the adiabatic Thouless protocol can be continuously deformed into a fine tuned finite-frequency cycle preserving the properties of noiseless quantized transport. 
    In this paper, we numerically investigate the implementation of such an ideal rapid-cycle Thouless pumping protocol in a one-dimensional optical lattice. It is shown that the rapidity will cause first order corrections due to next-to-nearest-neighbour hopping and second order corrections due to the addition of a harmonic potential.  Lastly, the quantization of the change in center of mass of the particle distribution is investigated, and shown to have corrections in the first order of the potential curvature.
\end{abstract}

\maketitle

\section{Introduction}
The past few decades have been marked by the discovery of various systems where topological properties of the quasiparticle spectrum are connected with the 
quantization of particle transport, for example Thouless pumping \cite{Thouless1983} or the integer quantum Hall effect \cite{Thouless1982, Niu1984}. In Thouless pumping, this integer valued particle transport is achieved by performing a non-contractible adiabatic loop through a non-degenerate parameter space. The amount of pumped charge can then be expressed by the Chern number associated with the Berry or Zak phase \cite{Berry1984, Zak1989, Xiao2010}. Although the original mathematics of the Thouless pumping dates back 30 years, the effect has only recently been observed directly using ultracold bosonic atoms in an optical superlattice \cite{Lohse2015, Nakajima2016}.

The adiabacity of the non-contractible loop is required to ensure the topological robustness of the Thouless pump. Generally, corrections to the quantization of particle transport arise when the parameter space is traversed at a finite frequency \cite{Wang2013, Privitera2018}. For special cases of Thouless pumping, such as parametric pumps \cite{Switkes1999, Brouwer1998, Altshuler1999, Levinson2001, Entin-Wohlman2002}, these non-adiabatic effects were studied \cite{Wang2013, Privitera2018, Ohkubo2008_1, Ohkubo2008_2, Cavaliere2009, Uchiyama2014, Watanabe2014}. In order to minimise corrections, strategies such as dissipation assisted pumping \cite{Arceci2020}, non-Hermitian Floquet engineering \cite{Hockendorf2020, Fedorova2020} and adiabatic shortcuts by external control \cite{Takahashi2020, Funo2020} were proposed. Recently however, a family of finite-frequency protocols on the Rice-Mele insulator have been constructed, in which all the quasi-excitations disappear altogether at the end of a rapid-cycle, resulting in a perfectly quantized and noise-free particle transport outside of the adiabatic limit \cite{Malikis2021}. Although this resolves the issue of non-adiabatic  breaking of topological quantization in an ideal homogeneous system, there might still be finite-size corrections \cite{Li2017} or corrections due to perturbations in the insulator, such as inhomogeneity due to an external potential. Understanding of such corrections is important in the context of
experimental realization of the rapid cycle pump, 
for example, in an ultracold atomic system.

In this paper, starting from the Rice-Mele insulator as the zeroth-order approximation, we investigate the corrections in the expectation value of the pumped charge w.r.t quantization due to performing a rapid-cycle protocol inside a one-dimensional optical lattice. Specifically, we investigate finite-size corrections, introduce next-to-nearest neighbour hopping and add a weak harmonic potential to the system. A lattice variant of the Weyl transform \cite{Case2008} is constructed to retrieve analytical relations between the corrections and the potential curvature. It is shown that all the corrections decay exponentially with the protocol defining parameters, which could also be chosen such that the corrections vanish completely. Lastly, a discussion is given on the change of center of mass after a rapid pumping cycle. It is shown that the corrections due to the rapid-cycle protocol under the harmonic potential are most pronounced in the change in center of mass, which is the current proposed and used method of measuring the charge pump \cite{Wang2013, Lohse2015, Nakajima2016}.
\section{Rapid-cycle Thouless pumping}
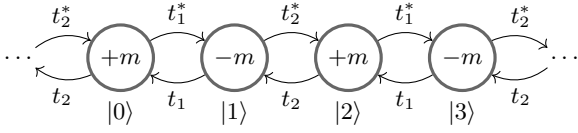
\begin{figure}
    \centering
    \begin{tikzpicture}[
        roundnode/.style={circle, draw=black!60, fill=white!100, very thick, minimum size=6mm},
    ]
        \node (X) {$\hdots$};
        \node[roundnode] (A) [right = 6mm of X] [label=below:$|0\rangle$] {$+m$};
        \node[roundnode] (B) [right = 6mm of A][label=below:$|1\rangle$]{$-m$};
        \node[roundnode] (C) [right = 6mm of B][label=below:$|2\rangle$]{$+m$};
        \node[roundnode] (D) [right = 6mm of C][label=below:$|3\rangle$]{$-m$};
        \node (Y) [right = 6mm of D]{$\hdots$};
        
        \draw[->, bend left] (X) to node [midway, above] {$t_2^*$} (A);
        \draw[->, bend left] (A) to node [midway, below] {$t_2$} (X);
        \draw[->, bend left] (A) to node [midway, above] {$t_1^*$} (B);
        \draw[->, bend left] (B) to node [midway, below] {$t_1$} (A);
        \draw[->, bend left] (B) to node [midway, above] {$t_2^*$} (C);
        \draw[->, bend left] (C) to node [midway, below] {$t_2$} (B);
        \draw[->, bend left] (C) to node [midway, above] {$t_1^*$} (D);
        \draw[->, bend left] (D) to node [midway, below] {$t_1$} (C);
        \draw[->, bend left] (D) to node [midway, above] {$t_2^*$} (Y);
        \draw[->, bend left] (Y) to node [midway, below] {$t_2$} (D);
    \end{tikzpicture}
    \caption{The tight-binding chain of the Rice-Mele model.}
    \label{fig:riceMeleChain}
\end{figure}

We begin with a recapitulation on the Rice-Mele model \cite{Rice1982}. This is a tight binding chain consisting of $2N$ atoms, on which there are orthonormal positional state $|\alpha \rangle$ which are subject to periodic boundary conditions, i.e. $|\alpha + 2N\rangle = |\alpha\rangle$. In the single-particle subspace, the Hamiltonian of this model is given by
\begin{equation}
    \begin{split}
        \hat H_{RM}(p) =  & \sum_{\alpha = 0}^{N-1}\bigg[ m\Big(|2\alpha\rangle \langle 2\alpha | - |2\alpha + 1 \rangle \langle 2\alpha + 1|\Big) \\
        &+ \Big(t_1|2\alpha\rangle \langle 2\alpha + 1| + t_2|2\alpha - 1\rangle\langle 2\alpha| + h.c.\Big)\bigg],
    \end{split}
    \label{eq:RMHamiltonian}
\end{equation}
where $p = (m, t_1, t_2)$ are the tight-binding parameters. A graphical representation of this model is shown in Fig.~\ref{fig:riceMeleChain}. The periodicity ensures that the Hamiltonian~(\ref{eq:RMHamiltonian}) can be written in the reciprocal space
\begin{equation}
    \hat H_{RM}(k, p) := \begin{bmatrix}
        m & t_1e^{\frac{ik}{2}} + t_2^*e^{-\frac{ik}{2}}\\
        t_1^*e^{-\frac{ik}{2}} + t_2e^{\frac{ik}{2}} & -m
    \end{bmatrix},
    \label{eq:RMHamiltonianReciprocal}
\end{equation}
where $k \in \mathscr{B} = \left\{-\pi + m\frac{2\pi}{N} \mid 0\leq m < N\right\}$, the discretised Brillouin zone. This Hamiltonian has two quasienergies $\epsilon_{\pm}(k, p)$ with the property $\epsilon_{-}(k, p) = -|\epsilon_+(k, p)|$. This creates two distinct energy bands which are seperated by the energy gap $E_{gap} = 2\sqrt{m^2 + \delta^2}$, where $\delta = |t_1| - |t_2|$. If we consider the parameter space $\mathscr{P} = \{(m,t_1, t_2) \mid E_{gap} > 0\}$ where this energy gap is strictly positive, then this space has a non-trivial fundamental group. Considering a non-contractible loop $p : [0, T) \rightarrow \mathscr{P}$ through this parameter space, we can look at the evolution of the lower energy Bloch states $|u_-(k,\tau)\rangle$ according to the Schr\"odinger equation
\begin{equation}
    \begin{split}
        i\frac{d}{d\tau} |u_-(k,\tau)\rangle = \hat H_{RM}(k, p(\tau))|u_-(k, \tau)\rangle \text{ with}\\
        \hat H_{RM}(k, p(0))|u_-(k,0)\rangle = \epsilon_-(k, p(0))|u_-(k,0)\rangle.
    \end{split}
\end{equation}
If the path through the parameter space is adiabatic, i.e. infinitely slowely, the Bloch states are ensured to be the lower eigenstates of the instantaneous Hamiltonian at all times and therefore no excitations will occur \cite{Born1928}. It can be shown that in the thermodynamic limit $N\rightarrow \infty$, the non-contractibility of the loop through the parameter space will then result in a non-zero pumped charge which is equal to the winding number of this loop around the degeneracy point \cite{Xiao2010}. This pumped charge can be directly related to the first Chern number associated with the Berry connection form \cite{Xiao2010, Berry1984}.

Outside of the adiabatic limit, i.e. at finite frequencies, there is generally a correction to this integer valued pumped charge \cite{Privitera2018}. Recently however, a family of protocols were constructed which results in noise-free integer valued Thouless pumping at finite frequencies \cite{Malikis2021}. Here, we investigate one such protocol. Consider the space $\{(x,y)\} = \R^2$ and a real oscillating function $\phi(y)$ which is a solution to
\begin{equation}
    \partial_y^2\phi + \sinh \phi = 0.
\end{equation}
The integrated form of this differential equation is given by
\begin{equation}
    (\partial_y\phi)^2 + 2\cosh \phi = 2\varepsilon,
\end{equation}
where $\varepsilon \in \R_{>0}$ and the period of $\phi$ will be denoted by $T_\phi(\epsilon)$. As the initial condition, we will choose $\phi(0) = 0$. It can be shown that this differential equation is equivalent to the zero curvature condition
\begin{equation}
    \partial_y \hat A_x - \partial_x \hat A_y + \left[\hat A_x, \hat A_y\right] = 0
    \label{eq:zeroCurvatureCond}
\end{equation}
for the anti-Hermitian matrix-valued vector fields
\begin{align}
    \hat A_x &= \frac{1}{4}\begin{bmatrix} i\partial_y \phi & 2\cosh\left(\frac{\phi - ik}{2}\right)\\
    -2\cosh\left(\frac{\phi + ik}{2}\right) & -i\partial_y \phi\end{bmatrix}\\
    \hat A_y &= \frac{i}{4}\begin{bmatrix}
        0 & 2\sinh\left(\frac{\phi - ik}{2}\right)\\
        2\sinh\left(\frac{\phi + ik}{2}\right) & 0
    \end{bmatrix}
\end{align}
where $k$ is a real valued parameter. The zero curvavature condition (\ref{eq:zeroCurvatureCond}) implies the existence of two orthonormal globally well-defined solutions $| F_{\pm}\rangle  \in \C^2$ of the system of equations
\begin{equation}
    \partial_x | F_{\pm}\rangle =\hat A_x|F_{\pm}\rangle, \quad \partial_y |F_{\pm}\rangle = \hat A_y |F_{\pm}\rangle.
    \label{eq:systemofequations}
\end{equation}
It should be noted that $|F_{\pm}\rangle$ depends on $x,y$ and $k$, which we will not write this down explicitly in the rest of this paper. Let $b\in \R$, $T = \frac{2\pi}{b}$ and the differentiable path $\gamma \colon \left[0, T\right] \rightarrow \R^2$ given by
\begin{equation}
    \gamma_x(\tau) = \tau, \quad \gamma_y(\tau) = \frac{T_\phi(\epsilon)}{2\pi}\left[b\tau - \sin(b\tau)\right],
    \label{eq:parameterPath}
\end{equation}
then we can define the matrix
\begin{equation}
    \hat H_\gamma(k, \tau) = i \dot{\gamma}_x\hat A_x + i\dot{\gamma}_y\hat A_y
    \label{eq:rapidRMHamiltonian}
\end{equation}
which coincides with the Rice-Mele Hamiltonian in reciprocal space (\ref{eq:RMHamiltonianReciprocal}). The solutions of the system of equations (\ref{eq:systemofequations}) will now evolve along the path $\gamma$ according to the Schr\"odinger equation
\begin{equation}
    i\frac{d}{d\tau} |F_{\pm}\rangle = \hat H_{\gamma} |F_{\pm}\rangle.
\end{equation}
Furthermore, since $\dot{\gamma}_y(0) = \dot{\gamma}_y\left(T\right) = 0$, the solutions $|F_\pm\rangle$ at $\tau = 0$ and at $\tau = T$ are the eigenstates of $\hat H_{\gamma}$, where we choose $|F_-\rangle$ to correspond to the lower eigenvalue. This protocol will result in a non-contractible loop in $\mathscr{P}$, such that the energy gap remains positive. This energy gap does however change during the evolution. Therefore, we will consider the function $$s(\tau) = \int_0^{\tau}d\tau'\  E_{gap}(\tau')$$ and reparametrize the path $\gamma$ in Eq.~(\ref{eq:parameterPath}) by
\begin{equation}
    \gamma(\tau) \mapsto \gamma
    \left(s^{-1}(\tau)\right) \text{ and } T \mapsto s(T)
    \label{eq:reparametrization}
\end{equation}
such that $E_{gap} = 1$ at all times. The non-contractibility of the loop in $\mathscr{P}$ and the fact that $|F_{-}\rangle$ is an eigenstate of the Hamiltonian at the start and end of the protocol will now result in a non-zero integer valued particle transport in the thermodynamic limit \cite{Malikis2021}.

Since this protocol works at finite frequencies, there are excitations of quasiparticles during the evolution. However, it makes sure that all of these excitations vanish at the end, such that the result after a rapid cycle is the exact same as with an adiabatic cycle. This is true for all values of $b$ and $\varepsilon$, which are the only two parameters the protocol depends on. The parameter $\varepsilon$ determines the width of the valence band and the conduction band. i.e.
\begin{equation}
    \max\{|\epsilon_\pm(k)|\} - \min\{|\epsilon_\pm(k))|\} = \frac{1}{2}\left[\sqrt{\frac{\varepsilon + 1}{\varepsilon - 1}} - 1\right],
\end{equation}
where it can be seen that the width of the bands becomes infinitely large in the limit $\varepsilon \downarrow 1$ and vanishes in the limit $\varepsilon \rightarrow \infty$. The parameter $b$ determines the steepness  and therefore the period of the path $\gamma$ as in Eq.~(\ref{eq:parameterPath}). A larger value for $b$ will also result in a more rapid pumping cycle. One should note that the angular frequency is in fact a function of both $b$ and $\varepsilon$, since the reparametrization (\ref{eq:reparametrization}) depends on $\varepsilon$. In the rest of this paper, we will investigate this specific protocol. It should be noted that other protocols could result in different specific properties. It is however expected that the general properties are similar for all rapid-cycle protocols.
\section{Finite-size corrections}
\begin{figure}
    \centering
    \includegraphics[width = 8.6cm]{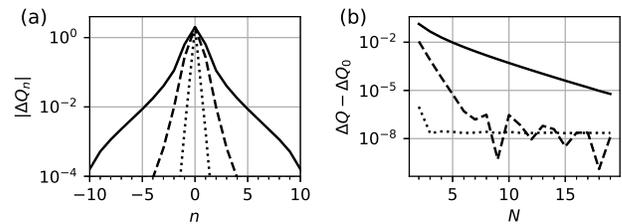}
    \caption{(a) Numerical calculation of the Fourier coefficients of the pumped charge for different values of $\varepsilon$. (b) Numerical calculation of the the finite-size corrections as function of $N$. Both plots have been made with $\varepsilon = 1.1$ (solid), $\varepsilon = 5$ (dashed) and $\varepsilon = 1000$ (dotted).}
    \label{fig:finiteCorrections}
\end{figure}

The proposed protocol gives a quantized particle transport outside of the adiabatic limit, but still only works in the thermodynamic limit. There are in general corrections to the pumped charge which decrease exponentially with $N$ \cite{Li2017}. Here, we will investigate those corrections for the rapid-cycle protocol specifically. If we consider the pumped charge per $k$-number, then because of its periodicity in $k$, it can be written as a Fourier series, i.e.
\begin{equation}
    \Delta Q(k) := \int_{0}^{T}\langle F_- | \partial_k \hat H_{\gamma} | F_-\rangle d\tau  = \sum_{n=-\infty}^{\infty} \Delta Q_n e^{ink},
\end{equation}
where $\Delta Q_n$ are the Fourier coefficients. In systems with a size $N\in \N$, the total pumped charge becomes
\begin{equation}
    \Delta Q := \frac{1}{N}\sum_{k \in \mathscr{B}}\Delta Q(k) = \Delta Q_0 + \sum_{m=1}^{\infty}\Delta Q_{mN} + \Delta Q_{-mN},
\end{equation}
which reduces to $\Delta Q = \Delta Q_0 \in \Z$ in the limit $N\rightarrow \infty$. Therefore, in finite-sized systems, the corrections are due to the additional Fourier-coefficients. Note that these finite sized corrections vanish if for all $n \in \N$, we have $\Delta Q_{n} = -\Delta Q_{-n}$ or $\Delta Q_n = \Delta Q_{-n} = 0$. It might be possible to construct a protocol in which this is true. In general however, this is not the case and there are still finite-sized correction. In the discussed rapid-cycle protocol, there is an analytical expression for the pumped charge per $k$-number. Namely, the pumped charge after one cycle due to the state $|F_{-}\rangle$ can be derived to be
\begin{equation}
    \begin{split}
        \Delta Q(k) = -\frac{1}{2}\int_0^{T_{\phi}(\epsilon)}\frac{\cosh(\phi(y)) + \cos(2k)}{\sqrt{2\epsilon + 2\cos(2k)}} dy.
    \end{split}
\end{equation}
Note that this is an even function, such that $\Delta Q_{-n} = \Delta Q_n$. Furthermore, it can be demonstrated that in the limit $\varepsilon \gg 1$ we get that $\Delta Q_n \ll 1$ for all $n \geq 2$. So in the limit where the width of the bands vanishes and the dispersion relation becomes flat, all finite-size corrections vanish for systems with $N \geq 2$. In Fig.~\ref{fig:finiteCorrections}(a), the Fourier coefficients have been plotted for different values of $\varepsilon$. It can indeed be seen that in the limit $\varepsilon \gg 1$, most Fourier coefficients are negligible. In Fig.~\ref{fig:finiteCorrections}(b), the finite size corrections are shown as function of $N$. It can be seen that as both $\varepsilon$ and $N$ increase, the finite sized correction start to vanish and become negligible w.r.t the numerical errors. Therefore, even for small systems it is possible to have a close to integer valued rapid-cycle Thouless pumping, where the corrections are actually independent on the rapidity.
\section{NN-hopping on the Optical superlattice}
\begin{figure}
    \centering
    \includegraphics[width = 8.6
    cm]{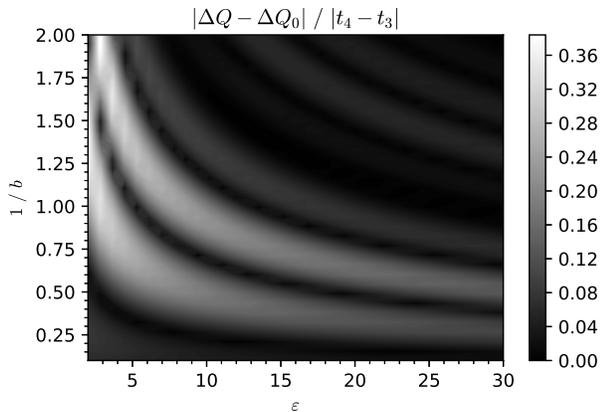}
    \caption{Numerical calculation of the corrections to integer valued pumped charge due to the NN-hopping terms as function of $\varepsilon$ and $1/b$.}
    \label{fig:ChargeNNHopping}
\end{figure}

Thouless pumping can be realized experimentally in a double-well optical superlattice of the form
\begin{equation}
    V(x,\tau) = -V_S(\tau)\cos^2\left(\frac{2\pi x}{d}\right) - V_L(\tau)\cos^2\left(\frac{\pi x}{d} - \phi(\tau)\right),
    \label{eq:opticalLattice}
\end{equation}
where $d$ is the lattice constant, $V_s$ and $V_L$ the depth of the short and long lattice respectively and $\phi$ the phase difference between the two lattices \cite{Wang2013, Lohse2015, Nakajima2016, Peil2003, Qian2011}. In the discussion of this lattice, we will use the unit of energy to be the recoil energy $E_R := \hbar^2 / (8md^2)$, where $m$ is the mass of the used atom. In the deep tight-binding limit, the two lowest energy bands of this model~(\ref{eq:opticalLattice}) can be approximated by those of the RM-Hamiltonian~(\ref{eq:RMHamiltonian}). Generally however, there is a slight difference between these two models. The band structure of the optical lattice can then be fully captured by considering higher hopping terms to the RM-Hamiltonian. Here, we will only consider the next-to-nearest-neighbour hopping terms, namely
\begin{equation}
    \begin{split}
        \hat H_{NN} =  &\sum_{\alpha = 0}^{N-1}t_3|2\alpha\rangle\langle 2\alpha + 2| + t_4|2\alpha - 1\rangle \langle 2\alpha + 1| + h.c.
    \end{split}
\end{equation}
which is added to the RM-Hamiltonian. With the addition of the NN-hopping terms, the two lower bands of the optical lattice coincide with those in the tight-binding approximation in sufficiently deep lattices, where the energy gap is much larger than the width of the bands. These extra NN-hopping terms will in general result in a deviation in the pumped charge. Specifically, when $|t_4 - t_3| \ll E_{gap}$, these corrections to integer values pumped charge are linearly dependent on the difference $|t_4 - t_3|$. For the ease of calculation, we will assume that the NN-hopping terms stay constant during the protocol. Although this is not generally true, this does give an idea of the order of magnitude, or at least an upper bound of the corrections due to the additional terms. In the rapid-cycle protocol, the deviation from integer valued pumped charge after one cycle is calculated as function of $\varepsilon$ and $1/b$ and shown in Fig.~\ref{fig:ChargeNNHopping}. It can be seen that the corrections are oscillatory in $1/b$ and $\varepsilon$, which means there are lines where the corrections vanish completely. Moreover, the amplitude of these oscillations decrease exponentially with both $\varepsilon$ and $1/b$. This means that in the adiabatic limit $b \rightarrow 0$ and in the limit of a flat dispersion $\varepsilon \rightarrow \infty$ there are no corrections to integer valued pumped charge due to NN-hopping terms.

\begin{figure}
    \centering
    \includegraphics[width = 8.6
    cm]{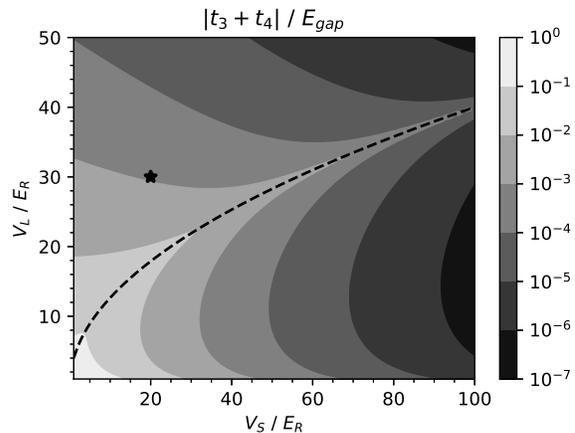}
    \caption{Numerical calculation of the NN-hopping terms per energy gap for the lower two bands of the optical lattice (\ref{eq:opticalLattice}) as function of $V_S$ and $V_L$ where $\phi= 0$. The dashed line is given by $V_S = V_L^2 / (16 E_R)$. The realization of Thouless pumping by Nakajima \textit{et al.} \cite{Nakajima2016} was done in an optical lattice with $(V_S, V_L) = (20, 30)E_R$, which is given by the star.}
    \label{fig:NNHopping}
\end{figure}

In order to calculate actual corrections, we should consider the magnitude of $|t_4 - t_3|$ in the optical lattice (\ref{eq:opticalLattice}). For simplicity however, we will only calculate the magnitude of the sum of the NN-hopping terms, i.e. $|t_3 + t_4|$, since this sum can be easily calculated by making use of the fact that 
\begin{equation}
    \epsilon_-(k) + \epsilon_+(k) = 2\cdot\text{Re}\left((t_3+t_4)e^{ik}\right),
\end{equation}
where $\epsilon_\pm(k)$ are the quasienergies of the RM-Hamiltonian with NN-hopping terms. In order to calculate the magnitude of the difference $|t_4 - t_3|$, one would need some fitting procedure for the bands. It is however expected that the order of magnitude of the difference $|t_4 - t_3|$ is similar to the order of magnitude of the sum $|t_3 + t_4|$. In Fig.~\ref{fig:NNHopping}, the magnitude of the sum of the NN-hopping terms per energy gap is plotted against $V_S$ and $V_L$ where $\phi = 0$. It can be seen that this magnitude decreases exponentially with both $V_S$ and $V_L$. Moreover, on the line $V_S = V_L^2 / (16 E_R)$, the NN-hopping terms are maximal, and this region should therefore be avoided to keep the NN-hopping terms to a minimum. As $\phi$ is varied, the absolute NN-hopping terms do not change significantly, while the energy gap does change. This will result in the ratio between the hopping constants and the energygap to change during the protocol, which already shows that the assumption that the NN-hopping terms stay constant is not true. The parameters $V_S$ an $V_L$ could also be varied during the protocol to overcome this problem.

Although the rapid-cycle protocol removes the finite-frequency corrections of adiabatic cycles, the NN-hopping terms introduce new corrections, whereas the topological quantization is still ensured in adiabatic cycles. Therefore, the parameters of the optical lattice should be chosen to minimize the NN-hopping terms. Also, to get the full characteristics of the optical lattice, even higher hopping terms should be considered. These are however expected to be negligible w.r.t the NN-hopping terms. Since we have not computed the exact mapping of the whole rapid-cycle protocol onto the optical lattice, we have not actually calculated the exact corrections that would occur in an optical lattice experiment, where there are most certainly varying NN-hopping terms. However, one should expect the order of magnitude of the corrections to be similar.

\section{Effect of Harmonic potential}
\label{sec:harmonicPotential}
\begin{figure}
    \centering
    \includegraphics[width = 8.6cm]{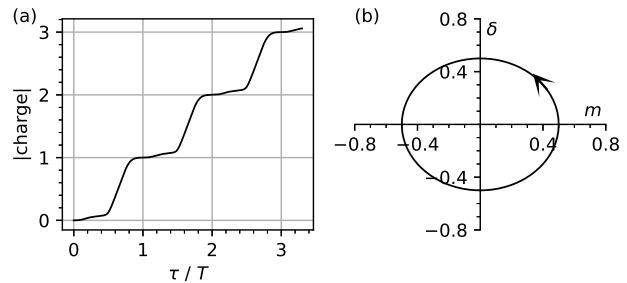}
    \caption{(a) Numerical calculation of the transported charge through the center of the lattice $\alpha_0$ in a system with $\xi = 0.005$. (b) The path taken through the parameter space $\{(m,\delta)\}$ in the rapid-cycle protocol with $b = 1$ and $\varepsilon = 2$.}
    \label{fig:rapidPumpPotential}
\end{figure}

In the optical lattice, the particles get trapped inside a harmonic potential, laid in the length of the lattice. In the single-particle subspace, this harmonic potential is given by
\begin{equation}
    \hat V = \sum_{\alpha = 0}^{N - 1}\frac{1}{2}\xi \left(\alpha - \alpha_0\right)^2\Big(|2\alpha\rangle \langle 2\alpha| + |2\alpha + 1\rangle \langle 2\alpha + 1|\Big),
    \label{eq:harmonicPotential}
\end{equation}
where $\xi \in \R_{>0}$ is analogous to the spring constant in a classical system, and $\alpha_0 = \frac{N-1}{2}$ is the center of the lattice. We will from now on consider $N$ to be odd, such that there is actually a center unit cell where the added potential vanishes. This added potential has the effect of localizing the eigenstates of the total Hamiltonian $\hat H = \hat H_{RM} + \hat V$. Here, we do not consider the NN-hopping terms, suppose the potential is smooth with $\xi \ll 1$ and the size of the system $N$ is large enough such that for the states
\begin{equation}
    \mathcal{S
    } = \{|\psi\rangle : \hat H |\psi\rangle = E|\psi\rangle \text{ and } E< 0\},
    \label{eq:states}
\end{equation}
the amplitude at the edges of the system become negligible. Here, $\mathcal{S}$ is the set of vacuum states in the zero temperature limit with a chemical potential $\mu = 0$. Using the fact that the potential is weak and smooth, the lattice looks locally unperturbed and periodic. Therefore, the non-contractible loop through $\mathscr{P}$ using the rapid-cycle protocol will still result in a non-zero particle transport. This particle transport is also close to integer as shown in Fig.~\ref{fig:rapidPumpPotential}, with some corrections due to the harmonic potential. To get these corrections, we will analyse the system in the phase space by introducing a lattice variant of the Weyl transform \cite{Case2008}. Namely, we define the Weyl transform of an operator $\hat A$ by the $2\times 2$ matrix $\tilde A(n,k)$ given by
\begin{equation}
    \begin{split}
        &\langle \alpha | \tilde A(n,k) | \beta\rangle = \\
        &\sum_{x = 0}^{N-1}e^{-ik\left(2x + \frac{\alpha - \beta}{2}\right)}\langle 2(n+x) + \alpha | \hat A | 2(n - x) + \beta \rangle
    \end{split}
    \label{eq:weylTransform}
\end{equation}
with $\alpha, \beta \in \{0,1\}$. More details and properties of this transformation are given in Appendix \ref{sec:WeylAppendix}. This Weyl transform can be applied to our system containing the Rice-Mele Hamiltonian (\ref{eq:RMHamiltonian}) and the added harmonic potential (\ref{eq:harmonicPotential}), such that the Weyl transform of the Hamiltonian is given by
\begin{equation}
    \tilde H(n,k)(\tau) = \hat H_{\gamma}(k, \tau) + \frac{1}{2}\xi n^2 \I_2,
    \label{eq:weylTransformedHamiltonian}
\end{equation}
which is the sum of the Rice-Mele Hamiltonian in reciprocal space (\ref{eq:rapidRMHamiltonian}) and a scalar matrix associated with the harmonic potential, translated such that the center of the lattice lies at $n = 0$. The Weyl tansform therefore simplifies to a two dimensional problem, in which we consider the Weyl transformed Liouvile-von Neumann equation
\begin{equation}
    i\frac{\partial \tilde \rho}{d\tau} = \widetilde{H\rho} - \widetilde{\rho H}.
    \label{eq:liouville-vonNeumann}
\end{equation}
As discussed in Appendix B, this gives rise to an expansion of the local vacuum density matrix in the insulating region, given by
\begin{equation}
    \tilde \rho(n) = \tilde \rho_0 + \xi\left(\tilde \rho_1 + n\tilde \rho_2\right) + \xi^2\left(\tilde \rho_3 + n\tilde \rho_4  + n^2 \tilde \rho_5\right) + \mathcal{O}\left(\xi^3\right),
    \label{eq:corrDensityMatrix}
\end{equation}
where $\tilde \rho_0$ is the local density matrix of the unperturbed lattice and the subsequent terms are corrections due to the harmonic potential. Eq.~(\ref{eq:corrDensityMatrix}) shows the general dependence of the density matrix on $n$ and $\xi$. Here, $\tilde \rho_1$ and $\tilde \rho_4$ are scalar matrices and also the only correction terms which have non-zero trace. This gives that in the limit $\varepsilon \gg 1$, the trace of the local density matrix is given by 
\begin{equation}
    \text{tr}(\tilde \rho(n)) = 1 + \frac{1}{4\varepsilon}\left(\xi + 3n\xi^2\right) + \mathcal{O}(\xi^3) > 1,
    \label{eq:traceDensity}
\end{equation}
meaning that inside the insulating region, the amount of particles per unit cell is greater than one and there is a slight cross-over with the conduction band in the vacuum state which scales inversely proportional to $\varepsilon$, as shown in Fig.~\ref{fig:densityDistribution}(b).
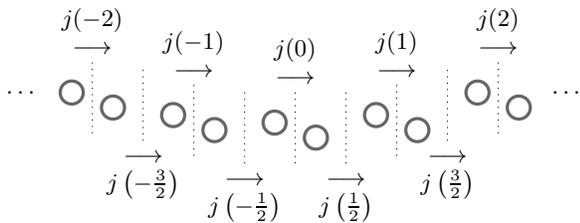
\begin{figure}
    \centering
    \begin{tikzpicture}[
        roundnode/.style={circle, draw=black!60, fill=white!100, very thick, minimum size=3mm},
    ]
        \node[roundnode] (E) {};
        \node[roundnode] (F) [right = 2mm of E, yshift = -2mm]{};
        
        \node (EJa) [right = 1mm of E, yshift = 4mm, align = center, anchor = south] {$j(0)$\\$\longrightarrow$};
        \node (EJb) [right = 1mm of E, yshift = -6mm, align = center, anchor = north] {};
        \draw[dotted] (EJa) -- (EJb);
        
        \node (EJc) [right = 7.75mm of E, yshift = 4mm, align = center, anchor = south]{};
        \node (EJd) [right = 7.75mm of E, yshift = -6mm, align = center, anchor = north] {$\longrightarrow$\\$j\left(\frac{1}{2}\right)$};
        \draw[dotted] (EJc) -- (EJd);

        \node[roundnode] (C) [left = 10mm of E, yshift = 1mm]{};
        \node[roundnode] (D) [right = 2mm of C, yshift = -2mm]{};
        
        \node (CJa) [right = 1mm of C, yshift = 4mm, align = center, anchor = south] {$j(-1)$\\$\longrightarrow$};
        \node (CJb) [right = 1mm of C, yshift = -6mm, align = center, anchor = north] {};
        \draw[dotted] (CJa) -- (CJb);
        
        \node (CJc) [right = 7.75mm of C, yshift = 3mm, align = center, anchor = south]{};
        \node (CJd) [right = 7.75mm of C, yshift = -7mm, align = center, anchor = north] {$\longrightarrow$\\$j\left(-\frac{1}{2}\right)$};
        \draw[dotted] (CJc) -- (CJd);
        
        \node[roundnode] (G) [right = 10mm of E, yshift = 1mm]{};
        \node[roundnode] (H) [right = 2mm of G, yshift = -2mm]{};
        
        \node (GJa) [right = 1mm of G, yshift = 4mm, align = center, anchor = south] {$j(1)$\\$\longrightarrow$};
        \node (GJb) [right = 1mm of G, yshift = -6mm, align = center, anchor = north] {};
        \draw[dotted] (GJa) -- (GJb);
        
        \node (GJc) [right = 7.75mm of G, yshift = 6mm, align = center, anchor = south]{};
        \node (GJd) [right = 7.75mm of G, yshift = -4mm, align = center, anchor = north] {$\longrightarrow$\\$j\left(\frac{3}{2}\right)$};
        \draw[dotted] (GJc) -- (GJd);
        
        \node[roundnode] (A) [left = 10mm of C, yshift = 3mm]{};
        \node[roundnode] (B) [right = 2mm of A, yshift = -2mm]{};
        
        \node (AJa) [right = 1mm of A, yshift = 4mm, align = center, anchor = south] {$j(-2)$\\$\longrightarrow$};
        \node (AJb) [right = 1mm of A, yshift = -6mm, align = center, anchor = north] {};
        \draw[dotted] (AJa) -- (AJb);
        
        \node (AJc) [right = 7.75mm of A, yshift = 3mm, align = center, anchor = south]{};
        \node (AJd) [right = 7.75mm of A, yshift = -7mm, align = center, anchor = north] {$\longrightarrow$\\$j\left(-\frac{3}{2}\right)$};
        \draw[dotted] (AJc) -- (AJd);
        
        \node[roundnode] (I) [right = 10mm of G, yshift = 3mm]{};
        \node[roundnode] (J) [right = 2mm of I, yshift = -2mm]{};
        
        \node (IJa) [right = 1mm of I, yshift = 4mm, align = center, anchor = south] {$j(2)$\\$\longrightarrow$};
        \node (IJb) [right = 1mm of I, yshift = -6mm, align = center, anchor = north] {};
        \draw[dotted] (IJa) -- (IJb);
        
        \node (X) [left = 2mm of A]{$\hdots$};
        \node (Y) [right = 7mm of I]{$\hdots$};
    \end{tikzpicture}
    \caption{The Rice-Mele chain under a harmonic potential. The current between atoms is calculated through the dashed lines as function of the unit cell.}
    \label{fig:riceMeleChainCurrent}
\end{figure}

We are now interested in the corrections to the total pumped charge due to the harmonic potential. The expansion of the density matrix~(\ref{eq:corrDensityMatrix}) gives the general dependence of the corrections in pumped charge on the position $n$ and the spring constant $\xi$. Here, it should be noted that $\tilde \rho_1$ and $\tilde \rho_4$ are scalar matrices and therefore have no contribution to the pumped charge. Moreover, since the position dependence is defined per unit cell, we have to distinguish between the current through a unit cell and between adjacent unit cells, as shown in Fig.~\ref{fig:riceMeleChainCurrent}. We can write the total pumped charge after one cycle through unit cell $n$ as
\begin{equation}
    \begin{split}
        \Delta Q(n) - \Delta Q_0 &= n\xi\big(A_1 + B_1\big)\\
        & + n^2\xi^2\big(A_2 + B_2\big)\\
        &+ \xi^2\big(A_3 + B_3\big) + \mathcal{O}(\xi^3)
    \end{split}
    \label{eq:pumpedChargeCorrections1}
\end{equation}
and the pumped charge between unit cells $n$ and $n+1$ as
\begin{equation}
    \begin{split}
        \Delta Q\left(n+\frac{1}{2}\right) - \Delta Q_0 &= \left(n+\frac{1}{2}\right)\xi\big(A_1 - B_1\big)\\
        &+\left(n+\frac{1}{2}\right)^2\xi^2\big(A_2 - B_2\big)\\
        &+ \xi^2 \big(A_3 - B_3\big) + \mathcal{O}(\xi^3),
    \end{split}
    \label{eq:pumpedChargeCorrections2}
\end{equation}
\begin{figure}
    \centering
    \includegraphics[width = 8.6cm]{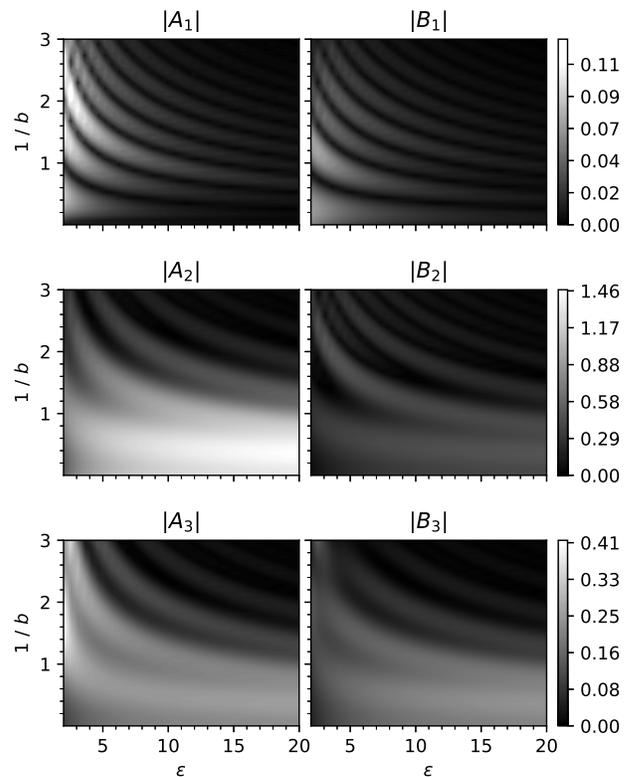}
    \caption{Numerical calculation of the pumped charge correction coefficients $A_i$ and $B_i$ as function of $\varepsilon$ and $1/b$. The $A_i$ terms represent the average pumped charge between each atom, while the $B_i$ terms describe the polarizing effect within the unit cells.}
    \label{fig:pumpedChargeCoefficients}
\end{figure}
where $\Delta Q_0 \in \Z$ is the unperturbed integer valued pumped charge and the terms $A_i$ and $B_i$ depend on the path through the parameter space, i.e. depend on $\varepsilon$ and $b$. Here, the $A_i$ terms can be thought of as the average pumped charge between each of the atoms, while the $B_i$ terms describe the polarization within the unit cells. In Fig.~\ref{fig:pumpedChargeCoefficients}, the numerical calculations of these coefficients are shown. It can be seen that that the behaviour of these functions is oscillatory, with minima where the coefficients become exactly equal to 0. The amplitude of these functions decays exponentially with both $\varepsilon$ and $1/b$. The non-zero $B_i$ coefficients cause a polarization in each unit cell, resulting in a change in energy. In the limit $\varepsilon \gg 1$, it can be derived that the the change in local energy is given by
\begin{equation}
    \begin{split}
        \Delta E(n) &= 2n\xi B_1 + n^2\xi^2\left(2B_2 - \frac{1}{2}(A_1 - B_1)\right)\\
                    &+ \xi^2\left(2B_3 - \frac{1}{4}\left(A_2 - B_2\right)\right) + \mathcal{O}(\xi^3).
    \end{split}
\end{equation}
This shows that in the rapid-cycle protocol, local excitations start to appear due to the harmonic potential. Interestingly, it possible to have a local correction to the pumped charge, while the expectation value of the local energy does not change. This suggests that there is additional noise on the energy and pumped charge, which could be investigated in further research.

Similar to the addition of NN-hopping, we can see that the rapid-cycle protocol in a harmonic potential creates additional corrections to an integer valued charge pump, whereas an adiabatic protocol only has finite-frequency corrections \cite{Privitera2018, Lohse2015, Nakajima2016}. At the center of the lattice, the corrections in a rapid-cycle scale with $\xi^2$, which makes them quite small for weak harmonic potentials, and could even be negligible w.r.t the correction due to NN-hopping in the optical lattice. When we go off-center, there are corrections which only scale linearly with $\xi$. However, since these corrections also scale linearly with $n$, the average pumped charge of a bulk around the center again scales quadratically with the potential curvature.

\section{Change in density distribution}
\begin{figure}
    \centering
    \includegraphics[width=8.6cm]{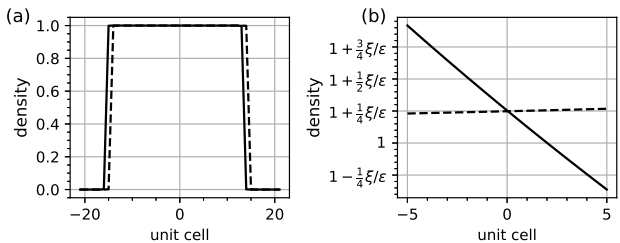}
    \caption{Numerical calculation of the density distribution per unit cell at $\tau = 0$ (dashed) and $\tau = T$ (solid) over the whole lattice (a) and zoomed in near the center of the lattice (b) for $\xi = 0.005$, $b = 1$ and $\varepsilon = 100$.}
    \label{fig:densityDistribution}
\end{figure}

In the optical lattice experiment, it is not actually the pumped charge through a point which is measured, but rather the change in center of mass of the whole density distribution \cite{Wang2013, Lohse2015, Nakajima2016}. As seen in Fig.~\ref{fig:densityDistribution}(a), the density distribution, and therefore also the center of mass, shifts after a pumping cycle. This measurement technique makes use of the fact that the pumping is close to integer inside the whole insulating region. However, as seen in Eq.~(\ref{eq:pumpedChargeCorrections1}) and Eq.~(\ref{eq:pumpedChargeCorrections2}), the pumped charge inside the insulating region depends on the position. By the continuity equation, this will result in a change in density distribution inside the insulating region after a pumping cycle. In Fig.~\ref{fig:densityDistribution}(b), this change in density distribution per unit cell is plotted after one cycle. Besides a change in density per unit cell, there is also the polarization of density in each unit cell. These two effects will result in a correction to the change in center of mass w.r.t integer value. Moreover, one should note that there are additional corrections due to the compressible region.

\begin{figure}
    \centering
    \includegraphics[width = 8.6cm]{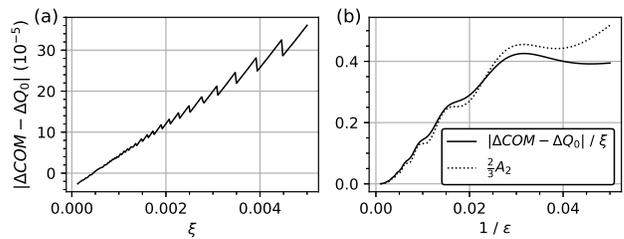}
    \caption{(a) The change in center of mass of the density distribution after a pumping cycle as function of $\xi$ with $b = 1$ and $\varepsilon = 100$. (b) The change in center of mass of the density distribution and $\frac{2}{3}A_2$ as function of $\varepsilon$ with $b = 1$ and $\xi = 0.005$.}
    \label{fig:errorCOM}
\end{figure}

As discussed at the end of Section~\ref{sec:harmonicPotential}, the average pumped charge in the bulk of the insulating region scales with $\xi^2$. It should be noted however, that the width of the insulating region also depends on $\xi$, and scales with $\xi^{-1/2}$. This will cause corrections to the change in center of mass to be linearly dependent on $\xi$. Specifically, using the continuity equation, Eq.~(\ref{eq:pumpedChargeCorrections1}) and Eq.~(\ref{eq:pumpedChargeCorrections2}), it can be demonstrated that the correction to the change in center of mass w.r.t integer value of the insulating region is given by
\begin{equation}
    \left|\Delta_{\rm COM} - \Delta Q_0\right| \approx \frac{2}{3}\xi A_2 + \mathcal{O}(\xi^2)
    \label{eq:errorCOM}
\end{equation}
in the limit $\varepsilon \gg 1$. In addition to the corrections due to the insulating region, the compressible region will also give some corrections. Although the compressible region is minimal in the same limit $\varepsilon \gg 1$, it does not vanish. In Fig.~\ref{fig:errorCOM}(a), it can be seen that the relation between the correction to the change in center of mass as function of $\xi$ is staggered. This can be explained by the fact that the width of both the insulating and compressible region is always integer valued. Therefore, a small variation in $\xi$ will not directly result in a variation in the width of the compressible region. When the variation in $\xi$ is large enough however, the compressible region will jump to the next atom, which causes the staggered behaviour. In Fig.~\ref{fig:errorCOM}(b), it can be seen that the Eq.~(\ref{eq:errorCOM}) gives a good approximation for $\varepsilon \gg 1$, where there are some slight corrections due to the compressible region.

The finite-frequency corrections to the pumped charge in adiabatic cycles scale with $\omega^2$ \cite{Privitera2018} and the potential corrections in a rapid-cycle protocol scale with $\xi^2$. However, the corrections to the change in center of mass scale only linearly with $\xi$. The rapid-cycle protocol might therefore introduce more corrections to the center of mass method than the adiabatic cycle would have given. Moreover, using this method, one also needs to take the corrections to the pumped charge due to the compressible region into account. Therefore, one might want to consider other methods in order to directly measure the actual pumped charge.
\section{Conclusions}
We have investigated various corrections that would occur for a specific rapid-cycle Thouless pumping protocol inside a one-dimensional optical lattice. Firstly, it was shown that the finite-sized corrections to an integer pumped charge decay exponentially with the size of the system, as seen in Fig.~\ref{fig:finiteCorrections}, and that these corrections vanish completely for systems with flat energy bands, such that the pumping is ideal even in systems of size 2.

Secondly, we gave some discussion on the order of magnitude of the corrections that would occur when we add NN-hopping terms to the RM-Hamiltonian (\ref{eq:RMHamiltonian}), which would occur in a realistic optical lattice. It was shown that these corrections vanish in the adiabatic limit, but that the rapidity of the cycle introduces new corrections, which are oscillatory and exponentially dependent on the rapidity and the width of the energy bands, and linearly dependent on the NN-hopping terms, as seen in Fig.~\ref{fig:ChargeNNHopping}.

Thirdly, we discussed the corrections with the addition of a harmonic potential to the band insulator. We constructed a lattice variant of the Weyl transform (\ref{eq:weylTransform}) to get the dependence of the corrections on the position and the potential curvature. These corrections can be split into an average effect and a polarizing effect, Eq.~(\ref{eq:pumpedChargeCorrections1}) and Eq.~(\ref{eq:pumpedChargeCorrections2}), which are both oscillatory and exponentially dependent on the rapidity and the width of the bands, as seen in Fig.~\ref{fig:pumpedChargeCoefficients}. Moreover, at the center of the lattice, the corrections to integer valued pumpd charge scale quadratically with the potential curvature.

Lastly, we gave a brief discussion on the change in center of mass of the particle distribution under the rapid-cycle protocol. Here, it was shown that the corrections in the change in center of mass are larger than the correction in actual pumped charge. Namely, these corrections are linear in the potential curvature. Moreover, the compressible regions also create additional corrections to the pumped charge, as seen in Fig.~\ref{fig:errorCOM}.

These investigated corrections give some insight for the realization of the rapid-cycle Thouless pumping protocol in an optical superlattice. It should be noted that this paper does not actually contain any numerical calculations of an optical superlattice, but rather discusses each correction separately. We have also not taken thermal effects or interactions between particles into account. The next step would therefore be to actually implement this rapid-cycle protocol onto an optical lattice.

\section{Acknowledgements}
This publication is part of the project Adiabatic Protocols in Extended Quantum Systems, Project No 680-91-130, which is funded by the Dutch Research Council (NWO). We would like to thank Savvas Malikis for helpful discussions.

\appendix
\section{The doubled-lattice Weyl transform}
\label{sec:WeylAppendix}
Here, we will give a discussion on the Weyl transform on a lattice with periodicity 2. Such a Weyl transform has actually already been constructed \cite{Fialkovsky2020}. However, it turned out to be not applicable to our system \cite{Buot2021}, and therefore a reformulation is needed. Here, we will only give a brief summary on this definition, and more specific details will be reported elsewhere.

We consider a doubled lattice of $N$ unit cells, where it is important that $N$ is odd valued. The intuitive reason for this is that we want a center unit cell, i.e. a 0 coordinate. When $N$ is considered to be even, this Weyl transform actually breaks down due to inconsistencies in the Fourier transform of the delta function. For a $2N\times 2N$ operator, the Weyl transform is now the $2\times 2$ matrix given by Eq.(\ref{eq:weylTransform}). Using the momentum basis, consisting of
\begin{equation}
    |k(\alpha)\rangle  = \frac{1}{\sqrt{N}}\sum_{n = 0}^{N-1} e^{\frac{ik(2n+\alpha)}{2}} |2n + \alpha\rangle 
\end{equation}
for $k\in \mathscr{B}$ and $\alpha \in \{0,1\}$, this can also be rewritten in the momentum basis, where the Weyl transform is given by
\begin{equation}
     \begin{split}
        &\langle \alpha | \tilde A(n,k) | \beta\rangle \\
        &= \sum_{p \in \mathscr{B}}e^{ip\left(2n + \frac{\alpha + \beta}{2}\right)}\langle (k+p)(\alpha)| \hat A | (k-p)(\beta) \rangle.
    \end{split}
    \label{eq:weylTransform2}
\end{equation}
Note the similarity with the one-dimensional and continuous Weyl transform \cite{Case2008}, where the biggest difference with the transformation given by Fialkovsky and Zubnov \cite{Fialkovsky2020} is that this transformation actually returns a matrix, just like a normal Fourier transformation on a doubled lattice would. The inverse of this transformation in the momentum basis is then given by
\begin{equation}
    \begin{split}
        &\langle p(\alpha) |\hat A | q(\beta) \rangle\\ &=\frac{1}{N}\sum_{n = 0}^{N-1}e^{-i\frac{p-q}{2}\left(2n + \frac{\alpha + \beta}{2}\right)} \langle \alpha | \tilde A\left(n, \frac{p+q}{2}\right) | \beta \rangle.
    \end{split}
    \label{eq:InverseWeylTransform2}
\end{equation}
A key property of the Weyl transform is that the trace of two operators $\hat A$ and $\hat B$ can be computed using the trace of the Weyl transforms, that is
\begin{equation}
    \text{tr}\left(\hat A \hat B\right) = \frac{1}{N}\sum_{n=0}^{N-1}\sum_{k\in \mathscr{B}} \text{tr}\left(\tilde A(n,k) \tilde B(n,k)\right).
    \label{eq:trace}
\end{equation}
Moreover, one can show that in the thermodynamic limit, the Weyl transform of the product of two operators $\hat A$ and $\hat B$ is given by
\begin{widetext}
\begin{equation}
    \langle \alpha|\widetilde{AB}(n,k) | \beta \rangle = \sum_{\gamma = 0}^{1}\langle \alpha |\tilde A(n,k)|\gamma \rangle e^{\frac{i}{2}\left[\overleftarrow\partial_n\left(\overrightarrow\partial_k - \frac{i}{2}(\beta - \gamma)\right) - \overrightarrow\partial_n\left(\overleftarrow\partial_k + \frac{i}{2}(\alpha - \gamma)\right)\right]}\langle \gamma | \tilde B(n,k) | \beta\rangle
\end{equation}
\end{widetext}
for $\alpha, \beta \in \{0,1\}$. Note the similarity with the Moyal product \cite{Fialkovsky2020}. The fact that $\tilde A$ and $\tilde B$ are matrices, will however result in additional correction terms in the exponentials. This form really only has meaning if the the exponent is expanded in a power series. We will add a formal parameter $\lambda$ to this expansion to keep track of the order in the expansion, which will be set to 1 later. The expansion of the Weyl transform of the product is then given by
\begin{widetext}
\begin{equation}
    \begin{split}
        \widetilde{AB}  &= \sum_{m = 0}^{\infty}\sum_{\alpha, \beta, \gamma = 0}^{1} \frac{\lambda^m}{m!} |\alpha\rangle \langle \alpha | \tilde A | \gamma \rangle \left[\frac{i}{2}\left[\overleftarrow\partial_n\left(\overrightarrow\partial_k - \frac{i}{2}(\beta - \gamma)\right) - \overrightarrow\partial_n\left(\overleftarrow\partial_k + \frac{i}{2}(\alpha - \gamma)\right)\right)\right]^m \langle \gamma | \tilde B | \beta\rangle\langle \beta|\\
        &=: \sum_{m = 0}^{\infty} \lambda^m f_m(\tilde A, \tilde B),
    \end{split}
    \label{eq:moyalexp}
\end{equation}
\end{widetext}
where we have introduced the functions $f_m(\tilde A, \tilde B)$, which are the $m$-th order expansion terms. This expansion of the product Weyl-transformation now gives rise to an expansion of the vacuum state of a perturbed doubled lattice, as we will show in the following section. This expansion has been shown to give correct predictions using numerical calculations, therefore suggesting that this definition of the Weyl transform is correct and useful. However, some more investigation on this transformation will be done and reported elsewhere.

\section{Expansion of the Weyl-transformed vacuum density matrix}
\label{sec:densityAppendix}
We will now consider the density matrix
\begin{equation}
    \hat \rho = \sum_{|\psi\rangle \in \mathcal{S}}|\psi\rangle \langle \psi| 
\end{equation}
with $\mathcal{S}$ as in Eq.(\ref{eq:states}). When we consider the unperturbed Rice-Mele chain, so when $\xi = 0$, this density matrix is just the sum over the outer-products of the Bloch-states of the lower band. The addition of a weak harmonic potential (\ref{eq:harmonicPotential}) with $\xi\ll 1$ will then result in small corrections to the density matrix. In particular, it will result in corrections to the Weyl transform of the density matrix. We can expand the Weyl transform of the density matrix according to the same formal parameter $\lambda$ as in Eq.(\ref{eq:moyalexp}), i.e.
\begin{equation}
    \tilde \rho(n,k) = \sum_{m=0}^{\infty} \lambda^m \tilde \rho_m(n,k),
\end{equation}
where $\tilde \rho_0 = |F_-\rangle \langle F_-|$, the vacuum density matrix of the Rice-Mele Hamiltonian (\ref{eq:rapidRMHamiltonian}). Importantly, the density matrix is idempotent, i.e. $\widetilde{\rho \rho} = \tilde \rho$. Therefore, it needs to satisfy the condition
\begin{equation}
    \label{eq:idemProp}
    \tilde \rho_m = \sum_{r+s+t = m}f_r(\tilde \rho_s, \tilde \rho_t).
\end{equation}
Moreover, it needs to commute with the Hamiltonian, i.e. $\widetilde{H\rho} - \widetilde{\rho H} = 0$. Defining $g_m(\tilde A, \tilde B) = f_m(\tilde A, \tilde B) - f_m(\tilde B, \tilde A)$ will then give the additional requirement
\begin{equation}
    [\tilde H, \tilde \rho_m] = -\sum_{\substack{r+s = m\\ s < m}}g_{r}(H, \tilde \rho_s).
    \label{eq:commProp}
\end{equation}
Finally, we can make use of the fact that
\begin{equation}
\label{eq:firstOrderProp}
    \tilde \rho_m\tilde \rho_0 + \tilde \rho_0 \tilde \rho_m = \tilde \rho_m + \frac{1}{\epsilon_-}\left(2\tilde \rho_m \hat H_\gamma + [\tilde H, \tilde \rho_m]\right)
\end{equation}
and combine it with Eq.(B3) and Eq.(B4) to get that the $m$-th order correction term in the Weyl transform of the density matrix is given by
\begin{equation}
    \tilde \rho_m = \frac{1}{2}\left[|\epsilon_-| \sum_{\substack{r+s+t = m\\ s,t < m}}f_r(\tilde \rho_s, \tilde \rho_t) + \sum_{\substack{r+s = m\\ s < m}}g_r(\tilde H, \tilde \rho_s)\right] \cdot \hat H_\gamma^{-1}
\end{equation}
which is a function of all the previous order correction terms, such that each correction term can be calculated through iteration. In the limit where $\xi \ll 1$ and $\varepsilon \gg 1$, the correction terms up to second order in $\xi$ can then be calculated to be
\begin{align}
    \begin{split}
        \tilde \rho_0 &= |F_-\rangle \langle F_-|
        \label{eq:zerothOrderTerm}
    \end{split},\\
    \begin{split}
        \tilde \rho_1 &= \frac{n\xi(1+e^{ik})}{\sqrt{16\varepsilon(1+\cos(k))}}|F_+\rangle \langle F_-| + h.c. + \mathcal{O}\left(\frac{1}{\varepsilon}\right)^{\frac{3}{2}}
    \end{split}\\
    \begin{split}
        \tilde \rho_2 &=  \frac{\xi}{8\varepsilon}\Big(|F_-\rangle \langle F_-| + |F_+\rangle \langle F_+|\Big)\\
        &+ \frac{n^2\xi^2}{8\varepsilon}\Big(|F_+\rangle \langle F_+| - |F_-\rangle \langle F_-|\Big)\\
        &+\left(\frac{n^2\xi^2(1+e^{ik})}{\sqrt{16\varepsilon(1+\cos(k))}}|F_+\rangle \langle F_-| + h.c.\right) +\mathcal{O}\left(\frac{1}{\varepsilon}\right)^{\frac{3}{2}},
        \label{eq:secondorderHom}
    \end{split}\\
    \begin{split}
        \tilde \rho_3 &= \frac{3n\xi^2}{8\varepsilon}\Big(|F_-\rangle \langle F_-| + |F_+\rangle \langle F_+|\Big) +\mathcal{O}\left(\frac{1}{\varepsilon}\right)^{2} + \mathcal{O}(\xi^3)
    \end{split}\\
    \begin{split}
        \tilde \rho_4 &= \frac{3\xi^2}{32\varepsilon}\Big(|F_+\rangle \langle F_+| - |F_-\rangle \langle F_-|\Big)
        \label{eq:fourthOrderHom}
    \end{split}\\
    \begin{split}
        \tilde \rho_m &= \mathcal{O}(\xi^3) \text{ for } m \geq 5.
        \label{eq:higherOrderTerm}
    \end{split}
\end{align}
It can be seen that this expansion results in an expansion of the density matrix in $\xi$. This shows the dependence of the local density matrix on $n$ and $\xi$ as given in Eq.(\ref{eq:corrDensityMatrix}), and gives the trace of the density matrix is given in Eq.~(\ref{eq:traceDensity})

\newpage
\bibliography{bibliography}

\onecolumngrid

\end{document}